\documentclass[longnamesfirst,apjl]{emulateapj}
\usepackage{apjfonts}

\newcommand{\beq}{\begin{equation}}
\newcommand{\eeq}{\end{equation}}
\newcommand{\bea}{\begin{eqnarray}}
\newcommand{\eea}{\end{eqnarray}}
\begin{document}

\title{On The Survival and Abundance of Disk-Dominated Galaxies}
\author{
Jun Koda\altaffilmark{1},
Milo\v s Milosavljevi\'c\altaffilmark{2}, 
and
Paul R. Shapiro\altaffilmark{2}
}
\altaffiltext{1}{Department of Physics, University of Texas, 1 University Station C1600, Austin, TX 78712.}
\altaffiltext{2}{Department of Astronomy, University of Texas, 1 University Station C1400, Austin, TX 78712.}

\righthead{THE ABUNDANCE OF DISK-DOMINATED GALAXIES}
\lefthead{KODA ET AL.}

\begin{abstract}

We study the formation of disk-dominated galaxies in a $\Lambda$CDM
universe.  Their existence is considered to be a challenge for the
$\Lambda$CDM cosmology, because galaxy mergers isotropize stellar
disks and trigger angular momentum transport in gas disks, thus
fostering the formation of central stellar spheroids. Here, we
postulate that the formation of stellar spheroids from gas-rich disks
is controlled by two parameters that characterize galaxy mergers, the
mass ratio of merging dark matter halos, and the virial velocity of
the larger merging halo.  We utilize merger histories generated from
realizations of the cosmological density field to calculate the
fraction of dark matter halos that have avoided spheroid formation,
and compare the derived statistics with the spheroid occupation
fractions in surveys of nearby galaxies.  We find, for example, that
the survival rate of disk-dominated galaxies in $\Lambda$CDM 
is just high enough to explain the observed fractional representation
of disk-dominated galaxies in the universe if the only mergers
which lead to central spheroid formation are those with mass ratios
$M_2/M_1>0.3$ and virial velocities $V_{\rm vir,1}>55\textrm{ km
  s}^{-1}$. We discuss the physical origin of this criterion.

\keywords{ cosmology: theory --- galaxies: formation --- galaxies: spiral  }

\end{abstract}

\section{Introduction}
\label{sec:intro}

The existence of disk-dominated galaxies, with little or no bulge, is
frequently cited as a challenge to the $\Lambda$CDM cosmology
\citep[e.g.,][]{Kautsch:06,Kormendy:07}. Apart from the question of
whether or not the theory of galaxy formation in $\Lambda$CDM can
succeed in making these observed rotationally-supported disk galaxies
in the first place \citep{D'Onghia:04, Abadi:03a}, the survival of
such disks, once formed, is our focus here.  The mergers that every
galaxy-hosting dark matter halo experiences can trigger angular
momentum transport in the interstellar medium of the merger remnant.
If a substantial amount of gas is transported into the central
kiloparsec of the remnant, the gas can fuel a starburst forming a
central stellar system---a ``classical'' bulge (i.e.,
self-gravitating, baryon-dominated stellar system supported by random motions rather
than rotation). By contrast, ``pseudobulges'' can also arise in some
disk galaxies if they have \textit{not} experienced recent major
mergers, by the secular transport of angular momentum \citep[e.g., by
galactic bars,][]{Jogee:05}.  These pseudobulges (sometimes called ``disky
bulges'') are supported more by rotation than random motion, however
\citep[e.g.,][and references therein]{Kormendy:04}.\footnote{
  Pseudobulges are to a larger degree supported by rotation than
  classical bulges and can have rotational velocity-to-1D velocity
  dispersion ratios $V_{\rm max}/\sigma > 1$.  Their velocity
  dispersions are smaller than expected from the Faber-Jackson
  relation.  Pseudobulges tend to have flattened isophotes and surface
  brightness profiles close to exponential. The
  classical bulges and pseudobulges can be distinguished from a third
  class, the boxy or peanut-shaped bulges, which are bars seen edge-on
  \citep{Bureau:99,Athanassoula:05}.}  About one third of all disk
galaxies in the local universe do not contain bulges or pseudobulges
\citep{Kautsch:06}\footnote{The classification of galaxies by their
  bulge content is usually done using simple bulge-disk or
  concentration models, which may not be accurate in extreme
  disk-dominated galaxies \citep[e.g.,][]{Boeker:03}.}  and another
third contain only pseudobulges.  \citet{Allen:06} carried out a
S\'ersic spheroid and exponential disk decomposition on a large sample
of galaxies and find that $30\%$ of exponential disks have small
bulge-to-total ratio $B/T<0.2$.  \citet{Barazza:07} report that $20\%$
of disk galaxies can be visually classified as bulgeless.\footnote{
  Pure disk galaxies also contain nuclear star clusters
  \citep[e.g.,][]{Boeker:02,Walcher:05} which could be products of
  secular gas transport \citep{Milosavljevic:04}, but these star
  clusters are tiny by comparison with the bulges described above.}

For our purposes here, we shall assume that the survival of
disk-dominance means that no classical bulge is formed.  Although the
precise characteristics of mergers that form classical bulges remain
unknown, disk-dominated galaxies must have avoided major mergers
during and after formation. Just how large the mass ratio of the
merging galactic halos must be in order to induce bulge formation is
somewhat uncertain, however. Mergers of similar-mass galaxies have
been shown to trigger starbursts and result in elliptical galaxies, by
gas dynamical and $N$-body simulations for galaxy halo masses $M
\gtrsim 10^{11} M_\odot$, merging at relative velocities of the order
of the virial velocity \citep{Mihos:96, Cox:07}. In that case, a bulge
forms from the momentum-exchange and compression of gas in merger
shocks and the outward angular momentum transport induced by merger
torques.  Pure $N$-body simulations of such mergers find that the
pre-existing stellar disks are mixed and destroyed
\citep[e.g.,][]{Naab:03}, leaving an elliptical galaxy. For minor
mergers, $N$-body and gas dynamical simulations with halo masses $M
\gtrsim 10^{11} M_\odot $ indicate that disks survive but bulges can
also grow \citep[][see also \citealt{D'Onghia:06}]{Mihos:94, Cox:07,
  Eliche-Moral:06}. However pure $N$-body simulations of this process
find that these bulges are pseudobulge-like (i.e., partially
supported by rotation), rather than classical, for mass ratios
$0.1\lesssim M_2/M_1\lesssim 0.25$ \citep{Bournaud:05}.  Bulge
formation by the merging of lower mass ($V_{\rm max} \lesssim
70\textrm{ km s}^{-1}$) and gas-rich galaxies has not yet been
simulated.


Semianalytic models of galaxy formation \citep[e.g.,][]{Kauffmann:93b,
  Baugh:96} assume they can track the morphological type of galaxies
by converting a disk component into a spheroidal component in mergers
with mass ratios $\mu\equiv M_2/M_1$ greater than some threshold.
These models successfully reproduce the distribution of earlier
morphological types by tuning the critical mass ratio for disk
destruction, and adopting a critical bulge-to-total mass ratio that
discriminates broadly between disk and elliptical galaxies.  Recent
semianalytic models employing merger trees extracted from hydrodynamic
$N$-body simulations yield a similar result \citep{Maller:06}.
However, those studies which focus on the survival of disks generally
do not address the abundance of {\it disk-dominated} galaxies. Existing ab
initio cosmological simulations have yielded disks with bulges
\citep[e.g.,][]{Abadi:03a}, but currently lack the dynamic range to
explore a large enough volume to sample the statistics of galaxy
morphology while simultaneously resolving the formation and mergers
of individual galaxies.

The purpose of this work is to compare the predicted disk survival
probabilities during hierarchical merging in a $\Lambda$CDM universe
with the observed statistics of galaxies at the end of the Hubble
sequence.  This comparison is then used to place constraints on the
physics of bulge-forming mergers.  In \S~\ref{sec:mergers}, we discuss
the role of mergers in bulge formation.  In \S~\ref{sec:jeans-mass},
we describe the effect of cosmic reionization and Jeans-mass filtering on
bulge formation. In \S~\ref{sec:trees}, we present a calculation of
bulge formation and disk survival probabilities derived from galactic
halo merger
trees generated from realizations of cosmological density fields.  In
\S~\ref{sec:results}, we compare observed disk galaxy statistics with
these merger tree results and place constraints on the properties of
bulge-forming mergers.  In \S~\ref{sec:discussion}, we summarize our
main conclusions.  Standard cosmological parameters consistent with
the {\it Wilkinson Microwave Anisotropy Probe} \citep{Spergel:07} are
assumed throughout.

\section{Disk Survival in $\Lambda$CDM Cosmology}
\label{sec:method}

\subsection{Mechanisms for Bulge Formation in Mergers}
\label{sec:mergers}

The merging of gas-rich galaxies fosters bulge formation directly and
indirectly.  Directly, the time-dependent gravitational potential of
the two merging components deflects some of the gas into the center of
the merger remnant, where it gets compressed in shocks and fuels a
starburst.  Indirectly, the gravitational tidal field of the merging
components excites nonaxisymmetric perturbations inside the merging
galaxies (bars, spirals, etc.) which then torque disk gas into the
center of the galaxy \citep[e.g.,][and references therein]{Combes:98}. 
In the center, again, shocks are
ubiquitous and play a role in angular momentum transport.  
The indirect channel
should be important in minor mergers, especially
where the smaller galaxy loses its gas to ram pressure 
stripping in the early stages of the merger.

The strength of direct merger torques is a function of the mass ratio
of the host dark matter halos of the merging galaxies, $\mu\equiv
M_2/M_1$.  The strength of nonaxisymmetric distortions in minor
mergers with $\mu\ll 1$ depends nontrivially on the resonance
structure of the merging halos and the orbit of the merger
\citep[e.g.,][]{Goldreich:80}.  These cannot be modeled in a general
case; therefore, we here consider only the gross properties of the
mergers, averaged over the merger orbital parameters and over the
properties of merging halos with given masses at a specific redshift.

Efficient  transport of angular momentum in perturbed gas disks
requires an excitation of nonlinear waves by 
nonaxisymmetric potential distortions that can give rise to 
momentum-transporting shocks.\footnote{For angular momentum 
transport by spiral shocks, see, e.g., \citet{Rozyczka:93}, 
\citet{Savonije:94}, and \citet{Goodman:01}.}  
The strength of merger shocks
is characterized by the Mach number  ${\cal M}_{\rm sh}$, which
is the ratio of the shock velocity $V_{\rm sh}$ to the sound speed of
the warm neutral gas $c_{\rm s}\sim 10\textrm{ km s}^{-1}$.  
Merger-driven strong shocks are radiative, because the
post-shock cooling time is much shorter than the dynamical and sound crossing times of the \ion{H}{1}
disk. In this limit the shocks are isothermal 
and the shock compression
is $\sim  {\cal M}_{\rm sh}^2$.  

In major mergers, we expect
$V_{\rm sh}\sim V_{\rm gal}$, where $V_{\rm gal}$ is the relative 
velocity of the two galaxies, while in minor mergers
$V_{\rm sh}\lesssim V_{\rm gal}$, 
although the forcing of the gas can be strong where
the gravitational torque is amplified locally by a resonance. 
In view of these
considerations, we postulate that, besides the mass ratio $\mu$, 
the efficiency of gas transport in mergers is 
controlled by a second parameter, the 
merger Mach number ${\cal M}_{\rm mer}\equiv V_{\rm vir,1}/c_{\rm s}\sim 
\beta^{-1} V_{\rm gal}/c_{\rm s}$, 
where $V_{\rm vir,1}$ is the virial velocity of the larger merging halo, 
and $\beta\sim 1$ is a dimensionless 
ratio of the orbital velocity $V_{\rm gal}$ of 
the two galaxies (i.e., merging dark matter halo centers) at the small 
radii that are relevant to bulge 
formation ($r\sim\textrm{few kpc}$) to the virial velocity 
$V_{\rm vir, 1}$.\footnote{Here and throughout the 
letter, $V_{\rm vir} \equiv
(GM_{200}/r_{200})^{1/2}$, where $M_{200}$ is the mass inside a
sphere with radius $r_{200}$ centered on the halo
within which the mean density equals $200$ times
the critical density of the universe.}

The relative velocity of gas disks could be smaller than the virial
velocity, i.e., $\beta<1$, if the merger starts affecting the gas only
when the distance between the halo centers has become much smaller
than the scale radii $r_{\rm s}$ in the \citet[][NFW]{Navarro:97}
profiles of the two halos.  The circular velocity of the NFW dark
matter density profile $\rho \propto (r/r_{\rm s})^{-1} (1+r/r_{\rm
  s})^{-2}$ at small radii $r \ll r_{\rm s}$ equals $V_{\rm circ}
\approx 1.3\ V_{\rm vir} (r/r_{\rm s})^{1/2}$, where the factor $1.3$
is for halo concentrations $c \equiv r_{200}/r_{\rm s} =10$.  The
factor depends only weakly on $c$; it is only slightly smaller ($\approx
1.1$) for $c=5$. The circular velocity $V_{\rm circ}$ reaches its
maximum at $r\approx2.2\ r_{\rm s}$. The scale radius equals
\begin{equation}
  r_{\rm s} = 75 \textrm{ kpc}\ c^{-1} \left(\frac{M}{10^{11} M_\odot}\right)^{1/3} 
[\Omega_{\rm m} (1+z)^3 + \Omega_\Lambda]^{-1/3} h^{-2/3} .
\end{equation} 
With this we find that 
\begin{equation}
\beta \approx (0.42+0.08\ z) \left(\frac{c}{10}\right)^{1/2}\left(\frac{r}{1\textrm{ kpc}}\right)^{1/2}
\left(\frac{M}{10^{11} M_\sun}\right)^{-1/6},
\end{equation}
implying that at low redshifts, the relative orbital velocity 
of the galaxy centers could be half the virial velocity of the larger merging halo 
when the separation is $r\sim 1\textrm{ kpc}$.

We emphasize, however, that in minor mergers, one must not attempt 
to identify ${\cal M}_{\rm sh}$ directly with $\beta {\cal M}_{\rm mer}$,
 because while ${\cal M}_{\rm mer}\gg 1$ in all cases of interest, 
the shock Mach number, which reflects the velocity of secondary gas
 flows in a perturbed merging system, can well be close to or below unity, 
implying a regime in which angular momentum transport is inefficient. 
The true condition for bulge formation triggering could be 
$\mu (\beta{\cal M}_{\rm mer})^\alpha\geq f_{\rm crit}$, where $f_{\rm crit}$ 
is a threshold and $\alpha$ is a power, e.g., $\alpha=1$ 
(the ``linear hypothesis'') and $\alpha=2$ (the ``compression 
ratio hypothesis'').

An important distinct possibility is that in which 
bulge formation is driven
\emph{cumulatively}, rather than 
induced in a single merger \citep{Bournaud:07}.  The central density 
of gas in the disk of a late-type disk galaxy could increase 
gradually due to slow, 
continuous radial gas inflow. Evidence for such inflow can be found in the ``central light excess'' (above the exponential law) 
in pure disk galaxies \citep[e.g.,][]{Boeker:03}. 
The inflow could be excited by perturbations 
associated with minor mergers.  Through their differential gravitational perturbations, many consecutive minor mergers can induce a slow, secular drift in the angular momentum distribution of the disk fluid, which could lead to central accumulation without leaving any characteristic signatures of merger-driven evolution (dynamically hot stellar components, etc.).   The inflow could also be driven by 
nongravitational processes, such as the magnetorotational instability in the gas disk
\citep[][see also, e.g., \citealt{Piontek:04}]{Milosavljevic:04}.  
The resulting increase of the central surface density brings the galaxy 
closer to the threshold for gravitational instability.  
The bulge or pseudobulge 
formation-triggering merger then must only nudge 
the galaxy over the threshold for, e.g., nuclear bar formation, 
where the galaxy is already marginally unstable.    

For systems which are gas-poor, collisionless mergers of stars can
result in an elliptical galaxy or a classical bulge, if and only if
the $\mu$ is sufficiently large \citep[e.g.,][]{Naab:03,
  Bournaud:05}. This bulge formation can be
characterized by merger ratio $\mu$, but it is independent of $V_{\rm
  vir,1}$ because the gravitational dynamics without gas is scale
free. However, there will still be a dependence on $V_{\rm vir, 1}$
through the requirement that the galaxies {\it prior} to their merger
were able to form long-lived (i.e. low-mass) stars (see below).

\subsection{The Critical Virial Velocity for Bulge Formation after Reionization
and Jeans-Mass Filtering}
\label{sec:jeans-mass}
In order for the merger of two halos to have
produced a bulge, the halos must have contained a substantial amount
of gas, or else stars already formed from collapsed gas. The gas content of small-mass halos, however, was affected by
the reheating of the intergalactic medium (IGM) out of which the gas inside
those halos collapsed, by cosmic reionization, a phenomenon known as
Jeans-mass filtering \citep{Shapiro:94}. The gas pressure of the
reheated intergalactic medium competes with gravitational instability,
in that case, 
to suppress structure formation in those baryons which would otherwise
have formed galaxies with virial velocity below some
threshold. The Jeans length in the IGM for a gas photoheated to $\sim
10^4 \textrm{ K}$ corresponds to a halo mass after collapse and
virialization for which the circular velocity is
\begin{equation}
  V_{\rm circ} = 55 \, (T_{\rm IGM} /10^4 \textrm{ K})^{1/2} \textrm{ km s}^{-1}
\end{equation}
\citep{Iliev:07}.  The actual threshold virial velocity is uncertain,
because one must account for the time-dependent growth of fluctuations
in an evolving background and because the formation of dark matter
halos affects the baryons in a nonlinear way. Estimates of the value
of the velocity threshold which results range from about $30$ to $80
\textrm{ km s}^{-1}$ \citep{Efstathiou:92, Thoul:96, Navarro:97b, Kitayama:00}. Whatever the precise value should be, this would
impose a lower limit to the critical virial velocity of merging halos
capable of producing a bulge, as described above, i.e., $V_{\rm
  crit,min} \sim 30 - 80 \textrm{ km s}^{-1}$ .

Since the virial velocity threshold which results from Jeans-mass
filtering depends primarily on the temperature, other sources of IGM
heating could have a similar effect. The supernova explosions
associated with massive star formation, for example, could also heat
the intergalactic gas. Such feedback could also have unbound the
interstellar gas from the galaxies which formed these stars, if the
galaxy virial velocities were small enough.

\subsection{Merger Histories of Low-Mass Galaxies}
\label{sec:trees} 

To explore the sensitivity of the fractional abundance 
of disk-dominated galaxies
produced during structure formation to the critical values of 
$\mu$ and $V_{\rm vir, 1}$, 
and thus to place constraints on the values
of these two parameters that are compatible with the observed statistics,
we generate merger histories of low-mass, disk-galaxy hosting halos
and study disk survival in this population of halos.  We utilize
those merger histories to calculate the abundance of disk-dominated
galaxies as a function of $\mu_{\rm crit}$ and 
$V_{\rm vir, crit}$.
We compare the resulting abundances with 
the incidence of late-type galaxies in the 
{\it Tully Galaxy Catalog} (\S~\ref{sec:results}).

The merger histories are generated from the nonlinear evolution of the
initial, linearly perturbed cosmological density field using the
publicly-available Lagrangian perturbation code PINOCCHIO
\citep{Monaco:02}.  The code generates a Gaussian-random field on a
cubic mesh, distributes particles on the mesh, and determines the
collapse time of each particle using an ellipsoidal collapse
criterion.  The ``collapsed'' particles are moved by Lagrangian
perturbation theory and related to virialized objects, which are the
dark matter halos, by a linking criterion.  We employed $512^3$
particles with cosmological parameters $\Omega_{\rm m}= 0.24$,
$\Omega_\Lambda= 0.76$, $\sigma_8=0.74$, and $h=0.73$, in a cubical
box with $50$ comoving Mpc on a side. The mass of an individual
particle was $m_{\rm part}=3.3\times 10^7 M_\odot$, and halos with
more than $10$ particles were selected for inclusion in the merger
tree.  For a given redshift, PINOCCHIO provides a list of all of the
halos with mass $M>10\ m_{\rm part}$ which formed inside the comoving
box at this or any {\it earlier} redshift, and a complete list of
their merger events. Each merger event is characterized by the merger
redshift and the masses of the halos participating in the merger.

We compute the fraction 
of halos containing disk-dominated galaxies as a function of the 
threshold for spheroid formation that is parametrized by 
the critical mass ratio $\mu$ 
and critical virial velocity $V_{\rm vir,1}$ of the larger halo
at the time of each merger. 
Specifically, we assume that a merger with 
$\mu>\mu_{\rm crit}$ will create a central stellar spheroid if the halo 
has a virial velocity
$V_{\rm vir,1}> V_{\rm vir,crit}$. 
We follow the most massive progenitor branch of the merger history of 
each halo and identify the resulting $z=0$ halo as containing a disk-dominated
galaxy if no spheroid has yet formed in the halo based on the defined criterion.

There are rare cases in which a progenitor mass is so small at high
redshift that bulge-forming mergers are not resolved by our numerical
results. We have estimated the number of such cases and confirmed that
it is negligible. For $1/10 < \mu_{\rm crit} < 1/3$, the fraction of
current halos in the mass range we will describe below, in which
bulge-forming mergers occur with a halo containing less than $50$ particles is
less than $3\%$. For $\mu_{\rm crit} = 1/2$, the fraction increases to
$8\%$, but the total mass of the merger remnant is small for those
mergers with $\mu>1/2$. If the smaller halo contains 50 particles, the
merger remnant with $\mu>1/2$ has at most 150 particles, which is ten
times smaller than $M_{\rm min}$, the minimum mass of interest for our comparisons with present-day galaxies.

We consider halos with present masses in the range $M_{\rm min}<M<
M_{\rm max}$, where $M_{\rm min}=5\times10^{10}M_\odot$ (corresponding
to $V_{\rm vir,min}\approx 60\textrm{ km s}^{-1}$) and $M_{\rm
  max}=10^{12}M_\odot$ (corresponding to $V_{\rm vir,max}\approx
160\textrm{ km s}^{-1}$); the resulting galaxy statistics are compared
with the observed galaxy statistics in the same approximate mass
range. [Since the halo mass and the maximum circular velocity of the
galaxy disk are not known for most of the galaxies in each observed
sample, we use the Tully-Fisher relation to estimate halo masses for
the observed galaxies.]  The present total number density of halos in
the above range is $0.021 \textrm{ Mpc}^{-3}$.  Disk-dominated galaxy
abundances thus calculated will not be strongly dependent on the
specific choice of $M_{\rm max}$ because halos with masses
$M<M_{\rm max}$ dominate the number density of halos in the universe today.
However, the abundances will be sensitive to $M_{\rm min}$. In
\S\ref{sec:results}, we explore the sensitivity to the choice of
$M_{\rm min}$.
An approach more accurate than the one employed here would dispense
with $M_{\rm min}$ and would consider halos of all masses and then
match the fractional disk and irregular galaxy abundances as a
function of halo mass.

We ignore the finite duration of the merger, which 
is the time elapsed between the halo contact and 
the final bulge formation. Indeed, the delay accounting for a
finite merger duration will affect only the low-redshift, 
disk-destroying mergers which are in the minority  
($z<1$, Fig. \ref{fig:redshift}). 
The dynamical friction time scale was recently  calibrated in 
$N$-body simulations \citep{Jiang:07},
\begin{equation}
  t_{\rm dyn} = \frac{0.94 \epsilon^{0.60} + 0.60}{0.86} 
               \frac{1}{\mu \ln(1+\mu^{-1})} \frac{r_{\rm vir}}{V_{\rm circ}} ,
\end{equation}
where $\epsilon$ is the circularity parameter. Setting
$\epsilon=0.5$ and $V_{\rm circ} = V_{200}$, this
simplifies to
\begin{equation}
  t_{\rm dyn} \sim \frac{0.14}{\mu} \frac{1}{\ln(1+\mu^{-1})} H(z)^{-1} ,
\end{equation}
where $H(z)^{-1}$ is the Hubble time at halo merger. If a pair of halos with
mass ratio $\mu>0.1$ merges before $z=1$, the galaxies in the 
halos merge
by $z=0$. In the halo mass range corresponding to disk-dominated galaxy hosts,
the present fraction of mergers in progress is only $\lesssim 5\%$ 
for mass ratios $\mu>0.2$ and $\lesssim 10\%$ for $\mu>0.1$.

\begin{figure}
  \plotone{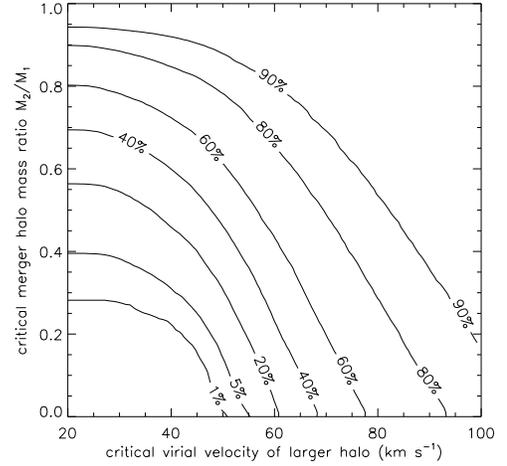}
  \caption{The fraction of disk-dominated galaxies (galaxies without
    classical bulges) that results from bulge formation criteria
    $(V_{\rm vir, crit}, \mu_{\rm crit})$
    characterized by the critical merger mass ratio, $\mu_{\rm crit}$,
    and the critical virial velocity of the larger halo at merger,
    $V_{\rm vir, crit}$.}
  \label{fig:diskfraction}
\end{figure}

\begin{figure}
\plotone{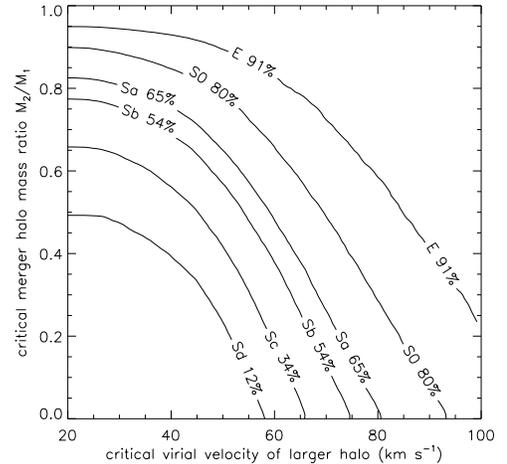}
\caption{Same as in Fig.~\ref{fig:diskfraction}, but the contours are
  labeled by the cumulative fraction of morphological types 
  in the Tully sample of galaxies.  Assuming that the type Sc and
  later do not contain classical bulges, the formation criteria along
  the Sc contour are compatible with the observed fraction of disk-dominated
  galaxies in the sample. }
\label{fig:tully}
\end{figure}


\subsection{Results}
\label{sec:results}

Figure~\ref{fig:diskfraction} shows the fraction of galaxies without
classical bulge as a function of the bulge formation criterion
$(V_{\rm vir,crit}, \mu_{\rm crit})$, which is the result of the model
described in \S~\ref{sec:trees}. In order to compare these theoretical
contours with the observed abundance of disk-dominated galaxies, we
select 2281 galaxies in the nearby universe from the {\it Tully Galaxy
  Catalog}\footnote{http://haydenplanetarium.org/universe/duguide/exgg\_tully.php}
that have blue magnitudes in the range $-20 < M_B < -17$ and are
located at distances $D<20\ h^{-1}\textrm{ Mpc}$ at which the catalog
is reasonably complete. This luminosity range is chosen to render the
number density of galaxies equal to the density $0.021 \textrm{
  Mpc}^{-3}$ of halos that we synthesize (see \S~\ref{sec:trees}).
The corresponding circular velocity range calculated from the
Tully-Fisher relation \citep[e.g.,][]{Kannappan:02} is $60 \textrm{ km
  s}^{-1} < V_{\rm c} < 160 \textrm{ km s}^{-1}$. This range is
consistent with the range of virial velocities in our theoretical halo
sample, which is a self-consistency check of our association of
galaxies with dark matter halos in our numerical halo catalog for
$\Lambda$CDM.

The assumption that the Tully-Fisher relation can be used to relate
the luminosities of galaxies in the Tully catalog to their halo virial
velocities (and, hence, to their masses, $M_{200}$) works best for the
spiral galaxies but is less certain for the elliptical and S0 galaxies.
The Tully-Fisher relation for S0 galaxies is shifted to lower
luminosity by about $M_B \sim +1.5$ for each $V_{\rm max}$, and the
scatter is larger compared to spirals \citep{Bedregal:06}. This
estimate is uncertain because the Tully-Fisher relation or virial
mass-to-light ratio is unknown for S0s at small mass near $M_{\rm
  min}$.  The virial mass-to-light ratio for elliptical galaxies could
be a factor of 10 larger relative to spirals in the $B$-band
\citep{Hoekstra:05, Guzik:02}.  If we shift the luminosity range of
the subsample that corresponds to $M_{\rm min}<M<M_{\rm max}$ by $+5$
mag for ellipticals and $+1.5$ mag for S0s, respectively, then all
the ellipticals will be removed (i.e., $M>M_{\rm max}$) and the number
of S0s will increase by $30\%$, but the sum of E and S0 will only
decrease from 25\% to 20\%, and the Sc contour move from 34\% to
36\%. Hence, uncertainties regarding the virial
velocities of the ellipticals and S0s in the sample do not
significantly affect our determination of the disk-dominated portion.

Figure \ref{fig:tully} is the same as Figure \ref{fig:diskfraction};
it shows the fraction of disk-dominated galaxies for various bulge
formation criteria.  Each pair of morphological type and fraction
printed on the contour indicates that that morphological type and
later types occupy the corresponding fraction in the subsample of the
{\it Tully} catalog.
If we choose to assume that a particular morphological type (e.g., Sd, Sc, or
Sb) and all later morphological types are disk-dominated galaxies,
while the earlier morphological types are galaxies with classical
bulges, then the fraction of disk-dominated galaxies in the subsample
is explained by the parameters $(V_{\rm vir,crit}, \mu_{\rm crit})$
along the contour labeled by the chosen transitional morphological type.  The
classification by morphological type does not precisely separate
galaxies that have classical bulges from those do not; nevertheless we
assume that Sc and later type galaxies are either bulgeless or have
pseudobulges, while Sbc and earlier types contain classical bulges.
Assuming this correspondence, the contour ``Sc'' in the figure shows
the parameter space locus yielding the abundance of galaxies without
classical bulges.



While the criterion for the survival in mergers of a given
morphological type must lie on the appropriate contour, from the
statistics alone it cannot be determined which specific value of
$(\mu_{\rm crit},V_{\rm vir,crit})$ along the contour is the true
physical criterion for bulge formation.  For example, 
the observed abundance of disk-dominated
galaxies is consistent with the hypothesis that mergers with $\mu>0.3$
in halos with $V_{\rm vir,1}>55\textrm{ km s}^{-1}$ create classical
bulges, while mergers that do not satisfy these criteria do not.  This
hypothesis is not unique; a somewhat larger $\mu_{\rm crit}$ and
somewhat smaller $V_{\rm vir,crit}$, and vice versa, would be equally
plausible on the basis of the statistics alone.
Were the critical velocity for bulge formation above $\sim~65 \textrm{
  km s}^{-1}$, however, then the relative abundance of disk-dominated
galaxies would be greater than observed for all mass ratios $\mu$. In
that case, there would be too few galaxies with bulges relative to
their observed abundance. A similar upper limit to $V_{\rm vir, crit}$
results from the fact that classical bulges are unlikely to form from
mergers that are too minor. If $V_{\rm vir, crit} \sim 65$, mergers
only produce enough disk-dominated galaxies if $\mu_{\rm crit} \ll 1$,
which may be implausibly small.  On the other hand, the critical
virial velocity cannot be much less than the minimum value 
imposed by Jeans-mass filtering discussed in
\S\ref{sec:jeans-mass}. This means that $V_{\rm vir,crit}$ cannot be much
\textit{less} than $\sim 60 \textrm{ km s}^{-1}$, either.  To identify
the true, unique criterion for bulge formation, one must resort to
physical insight to exclude implausible, extreme criteria that are
still allowed by the statistics.

If we
assume that $\mu_{\rm crit}$ lies between $0.4$ and $0.2$, then
the critical merger Mach number ${\cal M}_{\rm mer, crit} \equiv V_{\rm vir,crit}/(10 \textrm{ km s}^{-1})$ lies between $5$ and $6$.  When the
two halo centers have approached to within kiloparsecs of each other,
the true Mach number of the gravitational perturbation will be reduced
by the value of $\beta\sim \onehalf$.  Therefore, the question of the
physical plausibility of the criterion can be rephrased: In view of
the gravitational hydrodynamics of the gas disks in merging galaxies,
is it physically plausible that a merger with $\mu\sim
0.2-0.4$ and ${\cal M}_{\rm mer} \sim 5-6$ does not trigger
bulge formation, but a merger with a larger $\mu {\cal M}_{\rm
  mer}^\alpha$ does lead to central gas inflow and bulge formation?
For the ad hoc choice $\alpha=1$ and assuming $\beta\sim \onehalf$,
this yields a criterion $\mu_{\rm crit}(\beta{\cal M}_{\rm
  vir,crit})^\alpha\equiv f_{\rm crit}\sim 0.6-1$.

The \citet{Kautsch:06} sample of
edge-on disk galaxies gives a similar disk-dominated galaxy fraction.
If we assume the same additional fraction of early-type galaxies as in the Tully
subsample, the fraction of Sd is $11\%$ and the fraction of Sc or later
is $45\%$.


To test the dependence on mass cutoff $M_{\rm min}$, we vary this
parameter from the fiducial cutoff at $5\times10^{10}M_\odot$ to a
lower cutoff at $3.2 \times 10^{10} M_\odot$.  The number density of
halos increases by $50\%$ to $0.034 \textrm{ Mpc}^{-3}$.  To
compensate for the change in the number density of galaxies, we move
the lower luminosity cutoff for selection from the {\it Tully} sample
to $M_B<-15.5$.  The observed fraction of disk-dominated galaxies (Sc
or later morphological type) remains unchanged at $\approx 33\%$, but
the contours representing the bulge formation criterion move by about
$10\%$ toward lower critical virial velocities. This is because
mergers with a fixed mass ratio $\mu$ tend to occur at smaller $V_{\rm
  vir, 1}$ in smaller halos.  The criterion, e.g., with $(\mu_{\rm crit},V_{\rm
  vir, crit})= (\frac{1}{4}, 57 \textrm{ km s}^{-1})$ on the Sc
contour in Figure \ref{fig:tully} shifts only a small amount, to
$(\frac{1}{4}, 50 \textrm{ km s}^{-1})$, as the mass cutoff is lowered
to $3.2 \times 10^{10} M_\odot$. Any lower cutoffs than this are not
appropriate, because at luminosities $M_B\lesssim-15$, disk galaxies
give way to irregulars as the most common galaxy type
\citep[e.g.,][]{Binggeli:88}.

\begin{figure}
\plotone{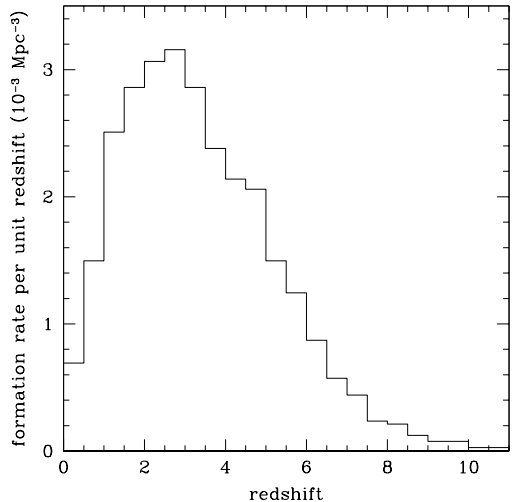}
\caption{The redshift distribution of first mergers that create
  classical bulges. We assumed $\mu_{\rm crit}=0.3$ and $V_{\rm
    vir,crit}=55 \textrm{ km s}^{-1}$ as the critical parameters for
  parameters of bulge formation, which is consistent with the
  abundance of disk-dominated galaxies in the local universe (see
  Fig. \ref{fig:tully}).}
\label{fig:bulgeformingz}
\end{figure}

\begin{figure}
\plotone{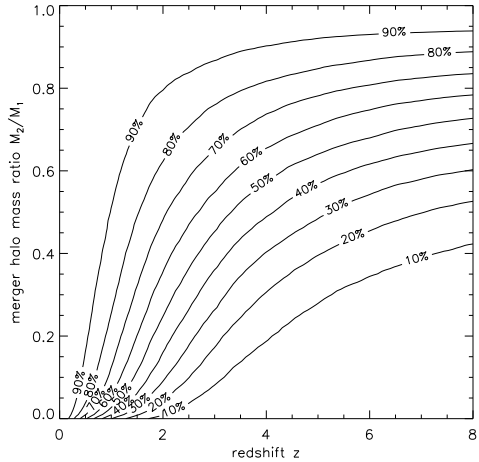}
\caption{Fraction of halos that have not experienced a merger with
  mass ratio $M_2/M_1$ or larger since redshift $z$.}
\label{fig:redshift}
\end{figure}

In Figure~\ref{fig:bulgeformingz}, we plot the redshift distribution
of the earliest bulge-forming mergers of $z=0$ halos for $\mu_{\rm
  crit}=0.3$ and $V_{\rm vir,crit}=55 \textrm{ km s}^{-1}$. The
distribution is insensitive to the choices of $\mu_{\rm crit}$ and
$V_{\rm vir,crit}$, as long as the two parameters remain on the same
contour in Figure \ref{fig:tully}.  This shows that bulge-forming
mergers generally took place long after reionization was completed
(i.e., $z_{\rm rei} \gtrsim 6$). Moreover, the typical collapse epoch
for these merging halos was at $z<6$ \citep{Shapiro:02}, so their star
formation was also post-reionization. Hence the lower limit to $V_{\rm
  vir, crit}$ imposed by Jeans-mass filtering is applicable.

In Figure \ref{fig:redshift}, we plot the fraction of halos that have
experienced a bulge-forming merger after redshift $z$.  The figure
shows that $60\%$ of the halos with masses in the range
$(0.5-10)\times10^{11}M_\odot$ have not experienced mergers with
$\mu\gtrsim 0.05$ after $z=1$, and $30\%$ have not experienced mergers
with $\mu>0.1$ after $z=2$.  \citet{Toth:92} placed constraints on the
mergers that could have taken place during the lifetime of a galactic
disk by quantifying the role of mergers in the heating and thickening
of the Milky Way's disk. The Galaxy could not have accreted more than
$5\%$ of its present mass during the past $5 \textrm{ Gyr}$, they
found. Subsequent work refined the estimates of disk heating,
resulting in less stringent constraints on the merger history
\citep[e.g.,][]{Velazquez:99, Benson:04}.  \citet{Kauffmann:93}
generated merger histories of Milky Way-sized halos using the
Press-Schechter excursion set theory and found that in an open
universe with $\Omega_{\rm m}=0.2$ and $\Omega_\Lambda=0$, the
abundance of disks is consistent with the Toth-Ostriker constraint. We
considered halos with masses smaller than that of the Milky Way, which
in the standard $\Lambda$CDM universe merge less frequently ($70\%$
have had no merger with $\mu>0.05$, and $80\%$ have had no merger with
$\mu>0.1$ since $z=1$) than the halos in earlier studies ($30\%$ for $\mu>0.05$
and $50\%$ for $\mu>0.1$ in \citealt{Kauffmann:93}).

\section{Conclusions}
\label{sec:discussion}
In order to explain the observed space density and fraction of
disk-dominated galaxies within the $\Lambda$CDM cosmology, we propose
a bulge-forming criterion such that only those halo mergers with mass
ratio greater than $\mu_{\rm crit} \sim 0.3$ and halo virial velocity (of the
larger halo) above $V_{\rm vir,crit} \sim 55 \textrm{ km s}^{-1}$
formed classical bulges, while other mergers did not. This criterion
has some degeneracy between $\mu_{\rm crit}$ and $V_{\rm vir,crit}$, but $V_{\rm
  vir,crit}$ cannot be larger than about $65 \textrm{ km s}^{-1}$
without underproducing the galaxy fraction with bulges, or much
smaller than $\sim 60 \textrm{ km s}^{-1}$ since Jeans-mass filtering
after reionization inhibits such small-mass halos from acquiring and
retaining baryons or forming stars.  This bulge-forming criterion also
gives a reasonable dimensionless condition, $\mu_{\rm crit} \beta
\mathcal{M}_{\rm mer,crit} \sim 1$, for the impact of merger on the
gas disk from the point of view of angular momentum transport. The
validity of this bulge formation criterion needs to be confirmed by
further analytic calculation or hydrodynamical simulations of mergers
in the halo mass range $V_{\rm vir,crit} \sim 60 \textrm{ km s}^{-1}$.

\acknowledgements

We would like to thank Shardha Jogee for detailed comments, and John Kormendy for inspiring and illuminating discussions.
This work was supported in part by NSF grant AST-0708795 to MM, and NASA ATP grants NNG04G177G and NNX07AH09G and NSF grant AST-0708176 to PRS.

\end{document}